\documentclass[epj]{webofc}
\usepackage[varg]{txfonts}   
\usepackage{bm}
\newcommand{\cA}{{\cal A}}  \newcommand{\cB}{{\cal B}}

  \newcommand{\cT}{{\cal T}}
  \newcommand{\cV}{{\cal V}}

\newcommand{\beq}{\begin{equation}} 
\newcommand{\eeq}{\end{equation}}
\newcommand{\beqa}{\begin{eqnarray}} 
\newcommand{\eeqa}{\end{eqnarray}}

\newcommand{\Tr}{{\textrm {Tr}}}

\newcommand{\cG}{{\mathcal G}}
\newcommand{\cH}{{\mathcal H}}
\newcommand{\cP}{{\mathcal P}}
\newcommand{\cW}{{\mathcal W}}

\newcommand{\cL}{{\cal L}}

\wocname{EPJ Web of Conferences}
\woctitle{ICNFP 2015}
\begin{document}
\selectlanguage{english}
\title{Anomalous transport in second order hydrodynamics~\thanks{Talk given by E.~Meg\'{\i}as at the 4th International Conference on New Frontiers in Physics (ICNFP 2015), 23-30 August 2015, Kolymbari, Crete, Greece.}}
%
%

\author{Eugenio Meg\'{\i}as\inst{1}\fnsep\thanks{\email{emegias@mppmu.mpg.de}} \and
        Manuel Valle\inst{2}
}

\institute{Max-Planck-Institut f\"ur Physik (Werner-Heisenberg-Institut), F\"ohringer Ring 6, D-80805, Munich, Germany
\and
        Departamento de F\'{\i}sica Te\'orica, Universidad del Pa\'{\i}s Vasco UPV/EHU, Apartado 644,  48080 Bilbao, Spain
}

\abstract{%
 We study the non-dissipative transport effects appearing at second order in the hydrodynamic expansion for a non-interacting gas of chiral fermions by using the partition function formalism. We discuss some features of the corresponding constitutive relations, derive the explicit expressions for the conductivities and compare with existing results in the literature.
}
\maketitle

\section{Introduction}
\label{sec:intro}

Hydrodynamics is an effective description of out-of-equilibrium systems in which it is assumed local thermodynamical equilibrium. The hydrodynamical systems should obey the conservation laws of the energy-momentum tensor and spin one currents, and these quantities are written in terms of fluid variables in the so-called constitutive relations. Some of the transport phenomena are related to dissipative effects, as they lead to entropy production: examples are the shear viscosity~$\eta$ and bulk viscosity~$\zeta$~\cite{Kovtun:2012rj}. However new phenomena on the hydrodynamics induced by quantum anomalies have recently received much attention and interest.  In presence of anomalies the currents are no longer conserved, and this has important effects in the constitutive relations. Two relevant phenomena appear at first order in the hydrodynamic expansion as a consequence of chiral anomalies: the {\it chiral magnetic effect}, which is responsible for the generation of an electric current induced by a magnetic field~\cite{Fukushima:2008xe}, and the {\it chiral vortical effect}, in which the electric current is induced by a vortex~\cite{Son:2009tf}. It is believed that these phenomena can produce observable effects in heavy ion physics~\cite{KerenZur:2010zw}, as well as in condensed matter systems~\cite{Basar:2013iaa}. These effects are non-dissipative, and the associated conductivities are almost completely fixed by imposing the requirement of zero entropy production. At second order a plethora of dissipative and non-dissipative conductivities have been studied, see e.g.~\cite{Kharzeev:2011ds} and references therein.

Some methods to compute the transport coefficients from a microscopic theory, either dissipative or non-dissipative, include kinetic theory~\cite{Arnold:2000dr,York:2008rr}, Kubo formulae~\cite{Landsteiner:2012kd}, diagrammatic methods~\cite{Manes:2012hf} and fluid/gravity correspondence~\cite{Bhattacharyya:2008jc}. Recently it has been proposed a new formalism to obtain the non-dissipative part of the anomalous constitutive relations, and it is based on the existence of an equilibrium partition function in a stationary background. It has been observed that the equations of hydrodynamics are significantly constrained by the requirement of consistency with the partition function~\cite{Banerjee:2012iz,Jensen:2012jh}, and these constraints seem to overlap with the ones obtained from the existence of an entropy current with non-negative divergence~\cite{Bhattacharyya:2014bha}.  In this work we study, within this formalism, the non-dissipative constitutive relations up to second order in the hydrodynamic expansion for an ideal gas of chiral fermions. We will be able to get explicit results for some of the transport coefficients.

\section{Hydrodynamics of relativistic fluids}
\label{sec:hydrodynamics}

Hydrodynamics is based on the assumption that the scales of variation of the observables are much longer than any microphysical scale in the system, and the result can be organized in a gradient expansion, also called hydrodynamic expansion~\cite{Kovtun:2012rj}. The constitutive relations for the energy-momentum tensor and charged currents write generally in the form
\begin{eqnarray}
  \langle T^{\mu\nu} \rangle &=& (\varepsilon + {\cal P}) u^\mu u^\nu  + {\cal P} g^{\mu\nu} + \langle T^{\mu\nu} \rangle_{\textrm{\scriptsize diss \& anom}}   \,,  \label{eq:T}  \\
\langle J^\mu \rangle &=& \rho u^\mu + \langle J^\mu \rangle_{\textrm{\scriptsize diss \& anom}} \,,  \label{eq:J}
\end{eqnarray}
where $\varepsilon$ is the energy density, ${\cal P}$ the pressure, $\rho$ the charge density and $u^\mu$ the local fluid velocity. In addition to the equilibrium contributions, there are extra terms which lead to dissipative and anomalous effects. Within the Landau frame~\footnote{This frame is defined as the one in which the viscous terms are transverse,  $u_\mu \langle T^{\mu\nu} \rangle_{\textrm{\scriptsize diss \& anom}} = 0 = u_\mu  \langle J^\mu \rangle_{\textrm{\scriptsize diss \& anom}}$.} and in presence of external electromagnetic fields, these terms write up to first order in derivatives as~\footnote{In this work we will consider $(3+1)$ space-time dimensions.}
\begin{eqnarray}
\langle T^{\mu\nu} \rangle_{\textrm{\scriptsize diss \& anom}} &=& - \eta P^{\mu\alpha}P^{\nu\beta} \left(\nabla_\alpha u_\beta + \nabla_\beta u_\alpha - \frac{2}{3}g_{\alpha\beta}\nabla^\lambda u_\lambda\right) - \zeta P^{\mu\nu} \nabla^\alpha u_\alpha + \cdots \,, \label{eq:T1} \\
\langle J^\mu \rangle_{\textrm{\scriptsize diss \& anom}}  &=& -\sigma T P^{\mu\nu} \nabla_\nu \left( \frac{\mu}{T} \right) + \sigma E^\mu + \sigma^\cB B^\mu + \sigma^\cV\omega^\mu + \cdots \,, \label{eq:J1}
\end{eqnarray}
where $P^{\mu\nu} = G^{\mu\nu} + u^\mu u^\nu$, and the electric, magnetic fields and vorticity are defined as $E^\mu = F^{\mu\nu} u_\nu$, $B^\mu = \frac{1}{2}\epsilon^{\mu\nu\rho\lambda}u_\nu F_{\rho\lambda}$ and $\omega^\mu = \epsilon^{\mu\nu\rho\lambda}u_\nu \nabla_\rho u_\lambda$ respectively.  The coefficients appearing in Eqs.~(\ref{eq:T1})-(\ref{eq:J1}) are the shear~$\eta$ and bulk~$\eta$ viscosities, the electric~$\sigma$, chiral magnetic~$\sigma^\cB$ and chiral vortical~$\sigma^\cV$ conductivities respectively. At this point it is worth analyzing the parity $\cP$ and time reversal $\cT$ properties of these coefficients. The spatial component of the charged current, $J^i$, is $\cP$-odd and $\cT$-odd, while the magnetic field and vorticity are $\cP$-even and $\cT$-odd. Then one concludes from Eq.~(\ref{eq:J1}) that $\sigma^\cB$ and $\sigma^\cV$  are $\cP$-odd and $\cT$-even. On the other hand, the second law of thermodynamics states the increase of entropy with time, i.e.
\begin{equation}
\frac{\partial}{\partial t} s  > 0    \,, 
\end{equation}
and this means that only $\cT$-odd contributions can lead to entropy production of the system. This is not the case of the chiral conductivities, so that these terms should be related to non-dissipative transport. A similar analysis for the shear, bulk and electric conductivities implies that these transport coefficients are $\cP$-even and $\cT$-odd, and they are associated with dissipative transport phenomena. In Sections~\ref{sec:eq_part_func} and \ref{sec:2ndOrder} we present the partition function formalism, which is suitable to compute $\cT$-even conductivities. We use this formalism to obtain the constitutive relations up to second order in the hydrodynamic expansion in Section~\ref{sec:const_rel}.

\section{Equilibrium partition function formalism to hydrodynamics}
\label{sec:eq_part_func}

We present in this section the main ingredients of the equilibrium partition function formalism introduced in Refs.~\cite{Banerjee:2012iz,Jensen:2012jh} (see also e.g. Refs~\cite{Manes:2013kka,Bhattacharyya:2013ida,Megias:2014mba,Bhattacharyya:2014bha}). Let us consider a relativistic invariant Quantum Field Theory with a time independent $U(1)$ gauge connection on the manifold
\begin{eqnarray}
ds^2 &=& G_{\mu\nu} dx^\mu dx^\nu = - e^{2\sigma(\bm{x})} (dt + a_i(\bm{x}) dx^i)^2 + g_{ij}(\bm{x}) dx^i dx^j \,,  \label{eq:G_background}\\
\cA &=& A_0(\bm{x}) dx^0 + \cA_i(\bm{x}) dx^i \,.  \label{eq:A_background}
\end{eqnarray}
The fields $\sigma$, $a_i$, $g_{ij}$, $A_0$ and $\cA_i$ are smooth functions of the spatial coordinates $\bm{x}$. As usual, the partition function of the system writes
\begin{equation}
Z = \Tr \, e^{-\frac{H-\mu_0 Q}{T_0}} \,,
\end{equation}
where $H$ is the Hamiltonian of the theory, $Q$ is the charge associated to the gauge connection, while $T_0$ and $\mu_0$ are the temperature and chemical potential at equilibrium. An obvious question is: what is the dependence of $Z$ on the fields $\sigma$, $a_i$, $g_{ij}$, $A_0$ and $\cA_i$? This has important implications for the hydrodynamics of the system, as we will see later. To answer this, the standard procedure is to build the most general partition function of the system consistent with the allowed symmetries, in particular: i) 3-dim diffeomorphism invariance; ii) Kaluza-Klein invariance, i.e. $t \to t + \phi(\bm{x}) \,, \; \bm{x} \to \bm{x}$; and iii) $U(1)$ time-independent gauge invariance (up to an anomaly). It is convenient to introduce the combination $A_i \equiv \cA_i - A_0 a_i$, which is invariant under the Kaluza-Klein transformation.

From the partition function we can compute the energy-momentum tensor and $U(1)$ charged current by performing the appropriate $t$-independent variations, i.e.
\begin{equation}
\delta \log Z = \frac{1}{T_0}\int d^3x \sqrt{g_3} \left( -\frac{1}{2} T_{\mu\nu} \delta g^{\mu\nu} + J^\mu \delta {\cal A}_\mu \right) \,,
\end{equation}
where $g_3 = \det (g_{ij})$. In particular, for a general partition function of the form $\log Z = \cW(e^\sigma, A_0, a_i, A_i, g^{ij},T_0,\mu_0)$, one gets
\begin{align}
\langle J^i \rangle &=\frac{T_0}{\sqrt{-G}}\frac{\delta \mathcal{W}}{\delta A_i} \,,  & 
\langle J_0 \rangle &=-\frac{T_0 e^{2\sigma}}{\sqrt{-G}}\frac{\delta \mathcal{W}}{\delta A_0} \,,  \label{eq:varJ}  \\
\langle T_0^{\;i} \rangle &=\frac{T_0}{\sqrt{-G}}\left(\frac{\delta \mathcal{W}}{\delta a_i} - A_0 \frac{\delta \cW}{\delta A_i}\right) \, ,&  \langle T_{00} \rangle &=-\frac{T_0 e^{2\sigma}}{\sqrt{-G}}\frac{\delta \cW}{\delta \sigma} \,, \label{eq:varT}
\end{align}
and a similar expression for $\langle T^{ij}\rangle$. This illustrates the fact that $\cW$ plays the role of a generating functional for the hydrodynamic constitutive relations, and one expects that its form matches order by other with the derivative expansion in hydrodynamics. In the rest of this section we present the most important properties of this partition function up to second order in derivatives.

\subsection{Equilibrium partition function at zeroth order}
\label{subsec:EPF0thorder}

We study first the zeroth order in derivatives. The most general partition function at this order, which is consistent with the symmetries mentioned above, reads~\cite{Banerjee:2012iz}
\begin{equation}
{\cal W}_{0} = \int d^3x \sqrt{g_3} \frac{e^\sigma}{T_0} P(e^{-\sigma}T_0, e^{-\sigma} A_0)  \,, \label{eq:W0}
\end{equation}
where $P$ is an arbitrary function of its arguments. From Eq.~(\ref{eq:W0}) and after applying the variational formulae~(\ref{eq:varJ})-(\ref{eq:varT}), one gets
\begin{eqnarray}
&& \langle J^0 \rangle = e^{-\sigma} \partial_b P \,, \qquad \langle J^i \rangle = 0 \,, \\
&&\langle T^{ij} \rangle = P \, g^{ij} \,,  \qquad \langle T_{00}\rangle = e^{2\sigma}(P - a \partial_a P - b \partial_b P) \,, \qquad \langle T_0^{\;i} \rangle = 0 \,,
\end{eqnarray}
where for convenience we have defined the variables $a \equiv e^{-\sigma} T_0$ and $b \equiv e^{-\sigma} A_0$. This result should correspond to the equilibrium contribution of the hydrodynamic constitutive relations. Finally, after a comparison with Eqs.~(\ref{eq:T})-(\ref{eq:J}) one gets
\begin{equation}
{\cal P} = P \,, \qquad \varepsilon = -P + a \partial_a P + b \partial_b P \,, \qquad \rho = \partial_b P \,, \qquad u^\mu = e^{-\sigma}(1, 0, 0, 0) \,.
\end{equation}
Note that $P$ is identified with the pressure of the system, and the thermodynamic variables $\varepsilon$, ${\cal P}$ and $\rho$ are determined in terms of this single {\it master} function. Thermodynamical consistency requires to identify the local value of the temperature, $T$, and chemical potential, $\mu$, with $a$ and $b$ respectively.

\subsection{Equilibrium partition function at higher orders}
\label{subsec:EPFhigherorder}

Let us discuss now the properties of the partition function at higher derivative orders. The most general partition function at first order in the derivative expansion was presented in~\cite{Banerjee:2012iz}, and it reads
\begin{equation}
\cW_1 =  \int d^3x \sqrt{g_3}  \left[ \alpha_1(T,\nu) \epsilon^{ijk}A_i F_{jk} + T_0\alpha_2(T,\nu) \epsilon^{ijk} A_i f_{jk} + T_0^2 \alpha_3(T,\nu) \epsilon^{ijk} a_i f_{jk}  \right] \,, \label{eq:W1}
\end{equation}
where 
\begin{equation}
 T = e^{-\sigma} \, T_0 \,, \qquad \nu = \frac{A_0}{T_0} \,,
\end{equation}
while $F_{ij} = \partial_i A_j - \partial_j A_i$ and $f_{ij} = \partial_i a_j - \partial_j a_i$. Eq.~(\ref{eq:W1}) contains only $\cP$-odd contributions, as one cannot build $\cP$-even contributions to the partition function at first order in derivatives. Following the method of Section~\ref{sec:2ndOrder}, a computation of the partition function for an ideal gas of Weyl fermions leads to the following result~\cite{Megias:2014mba}
\begin{equation}
\alpha_1(T,\nu) =  -\frac{C}{6} \nu \,, \quad \alpha_2(T,\nu) =  -\frac{1}{2}\left( \frac{C}{6}\nu^2 - C_2 \right) \,, \quad \alpha_3(T,\nu) = 0 \,,
\end{equation}
where 
\begin{equation}
C = \frac{1}{4\pi^2} \,, \qquad C_2 = \frac{1}{24}  \,, \label{eq:CC2}
\end{equation}
are constants related to the axial anomaly~\cite{Son:2009tf,Erdmenger:2008rm} and gauge-gravitational anomaly~\cite{Landsteiner:2011cp} respectively. After a suitable computation of the constitutive relations as it will be explained in Section~\ref{sec:const_rel}, one gets the well known expressions for the chiral conductivities in the Landau frame~\footnote{We have considered a fluid with an anomalous charge $U(1)$. The extension to $U_V(1) \times U_A(1)$ would require the introduction of both vector and axial gauge connections.}
\begin{equation}
\sigma^\cB = C \mu -\frac{\rho}{\varepsilon + {\cal P}} \left( \frac{1}{2}C\mu^2 + C_2 T^2 \right) \,, \qquad \sigma^\cV = \frac{1}{2} C \mu^2 + C_2 T^2  - \frac{\rho}{\varepsilon + {\cal P}} \left( \frac{1}{3}C\mu^2 + 2C_2 T^2 \right) \mu \,, \label{eq:sigmaBV1storder}
\end{equation}
where we have used that $\nu=\mu/T$. These results have been obtained in a wide variety of methods, see e.g.~\cite{Fukushima:2008xe,Erdmenger:2008rm,Son:2009tf,Banerjee:2008th,Landsteiner:2011cp,Landsteiner:2011iq,Landsteiner:2012kd,Jensen:2012kj,Megias:2013xla,Megias:2013joa,Landsteiner:2013aba}.

Finally, the most general partition function at second order in derivatives is built from seven scalar and two pseudo-scalar quantities as follows~\cite{Bhattacharyya:2013ida,Bhattacharyya:2014bha}
\begin{equation}
\begin{split}
\cW_2 &= \int d^3 x \sqrt{g_3} \Bigl[ M_1 g^{i j} \partial_i T \partial_j T + 
  M_2 g^{i j} \partial_i \nu \partial_j \nu  + 
  M_3 g^{i j} \partial_i \nu \partial_j T    \\ 
  &\quad +T_0^2 M_4 f_{i j} f^{i j} +    M_5 F_{i j} F^{i j} + T_0 M_6 f_{i j} F^{i j} + M_7 R \\
 &\quad + N_1 \epsilon^{ijk} \partial_i A_0 f_{jk} + T_0^{-1} N_2 \epsilon^{ijk} \partial_i A_0 F_{jk}  {\Bigr]}  \,,   
\end{split}  \label{eq:W2}
\end{equation}
where $R$ is the Ricci scalar in 3 dim, with $M_i = M_i(T,\nu)$ and $N_i = N_i(T,\nu)$. We keep for the moment both $\cP$-odd and $\cP$-even contributions.

\section{Second order partition function}
\label{sec:2ndOrder}

We present in this section the procedure to compute the second order partition function of Eq.~(\ref{eq:W2}) for a theory of free massless Dirac fermions. We refer to Ref.~\cite{Megias:2014mba} for full details in the computation.

\subsection{Free theory of Dirac fermions}

We will derive the partition function by using Pauli-Villars (PV) regularization, and this demands the consideration of the massive theory for the vacuum contribution. The action of the theory is
\begin{equation}
S = \int d^4x \sqrt{-G} \cL \,, \qquad \textrm{where} \qquad \cL = -i\bar\Psi\underline\gamma^\mu\nabla_\mu\Psi +im\bar\Psi\Psi \,.
\end{equation}
The space-time dependent Dirac matrices satisfy~$\{\underline\gamma^\mu(x), \underline\gamma^\nu(x)\}=2G^{\mu\nu}(x)$, and they are related to the Minkowski matrices by $\underline\gamma^\mu(x)=e^\mu_a(x) \gamma^a$, where $e_a^\mu(x)$ is the vierbein, $\{ \gamma^a, \gamma^b\} = 2\eta^{ab}$ and $\eta^{a b} = \text{diag}(-1,1,1,1)$. The $U(1)$ current and energy-momentum tensor write
\begin{equation}
J^\mu = -\bar\Psi\underline\gamma^\mu  \Psi  \,,  \qquad T_{\mu \nu} =  \frac{i}{4} \bar{\Psi} \left[ 
\underline\gamma_\mu \overrightarrow{\nabla}_\nu  - \overleftarrow{\nabla}_\nu \underline\gamma_\mu +  ( \mu \leftrightarrow \nu) \right ] \Psi \,, 
\end{equation}
where it has been assumed that the spinor field $\Psi = \left( \begin{array}{c} 
\psi_L \\
\psi_R  \\
\end{array} \right)$  satisfies the Dirac equation. Using the explicit form of the background Eqs.~(\ref{eq:G_background})-(\ref{eq:A_background}) one has
\begin{align}
J_0 &= - e^{-\sigma} \psi^\dagger \psi \,, \qquad J^i = - \psi^\dagger \sigma_i \psi \,,  \label{eq:J_L} \\
T_{00} &= \frac{i}{2} e^{\sigma} \left( \psi^\dagger \partial_t \psi - \partial_t\psi^\dagger \psi \right) 
+ e^\sigma A_0 \psi^\dagger \psi 
- \frac{1}{4} e^{3\sigma} \epsilon^{ijk} \partial_j a_k \psi^\dagger {\bf \sigma}_i \psi \,,  \label{eq:T00_L}
\end{align}
where $\sigma_i$ are the Pauli matrices, and $\psi$ is the two-component Weyl fermion $\psi_L$. The same expressions are obtained for $\psi_R$, but with opposite sign in $J^i$.

\subsection{Thermal Green's function}
\label{subsec:Green}

The expectation values of $J_\mu$ and $T^{\mu\nu}$ at equilibrium may be computed from the thermal Green's function, defined as
\begin{equation}
\langle T \psi(-i\tau,  \bm  x)\psi^\dagger(0, \bm  x') \rangle_{T_0} =
T_0\sum_n e^{-i\omega_n\tau} \mathcal{G}( \bm  x,  \bm  x',\omega_n)\,, \quad \omega_n = 2\pi T_0\left( n + \frac{1}{2}\right) \,, \label{eq:G_def}
\end{equation}
where $T$ denotes time ordering. The explicit calculation of the Green's function is the most computationally demanding step in the derivation. After rotating to imaginary time $t \to -i\tau$, the Green's function satisfies
\begin{equation}
-\sqrt{-G}\,\gamma^0 \underline\gamma^0(i\omega_n- \cH)\cG (\bm{x},  \bm{x}^\prime,\omega_n)= \delta^{(3)}(\bm{x}- \bm{x}^\prime) \,, \label{eq:eqG}
\end{equation}
where $\cH$ is the Hamiltonian
\begin{equation}
\cH = -i\left (\frac{1}{4} \omega_0^{\;\; a  b}\gamma_{ab} - i A_0 \right) -
\frac{i}{g^{00}}\, \underline\gamma^0  \left( \underline\gamma^k\,\nabla_k - m \right) \,,
\end{equation}
with $\omega_\mu^{\;\; a b}$ the spin connection and $\gamma_{ab} = \frac{1}{2}[\gamma_a, \gamma_b]$. Then by expanding Eq.~(\ref{eq:eqG}) in derivatives,  the Green's function can be computed recursively order by order in a derivative expansion, i.e. $\cG = \cG_0 + \cG_1 + \cG_2 + \cdots$, where $\cG_\ell$ is the contribution at order $\ell$. Note that to get $\cG_\ell$ one needs to know the Green's function at all the orders lower than $\ell$. The complete expressions for $\cG$ up to second order are very lengthy and will not be presented here. Finally, from Eqs.~(\ref{eq:J_L})-(\ref{eq:G_def}) one obtains the precise form of the current and energy-momentum tensor up to $\ell$-th order in the derivative expansion
\begin{align}
\langle J_0 \rangle_\ell &= T_0\sum_n \left[-e^\sigma \mathrm{tr}\, \cG_\ell(\bm  x, \bm x, \omega_n)\right] \,,  \qquad \langle J^i \rangle_\ell = -T_0\sum_n \mathrm{tr}\left[ \sigma_i\,\cG_\ell(\bm  x, \bm  x, \omega_n)\right] \,,  \label{eq:Jp} \\
\langle T_{00} \rangle_\ell &= T_0 \sum_n \left[ e^\sigma (i \omega_n + A_0)\,  \text{tr}\, \cG_\ell(\bm  x, \bm  x, \omega_n) - 
    \frac{1}{4} e^{3 \sigma}  \epsilon^{ijk} \partial_j a_k\,\text{tr}\left[\sigma_i\, \cG_{\ell-1}(\bm  x,\bm  x, \omega_n)\right]  \right]\,.  \label{eq:Tp}
\end{align}
Note that $\langle T_{00} \rangle_\ell$ receives a contribution $\propto \text{rot}\,  \bm{a} \cdot \langle \bm{J} \rangle_{\ell-1}$.

\subsection{Charge density at second order}
\label{subsec:charge_density}

To get ${\cal W}_2$ it is enough to compute $\langle J_0\rangle_2$ and $\langle T_{00}\rangle_2$ including only bilinear terms $\sim \partial_i X \partial_j Y$. The evaluation of Eq.~(\ref{eq:Jp}) produces the following renormalized expression 
\begin{align}
\langle J_0\rangle_2 &=  \frac{1}{24 \pi^2} \biggl(
  -\nabla^i A_0 \nabla_i \sigma + \frac{1}{2}e^{2 \sigma}  f_{i j} F^{ij}  + 
  \frac{1}{2} A_0 e^{2 \sigma}f_{i j} f^{ij} \biggr) \left(\ln \frac{e^{2 \sigma} \bar{M}^2}{T_0^2} + Q (\nu)\right)  \notag \\ 
& \quad  + \frac{1}{48 \pi^2} \left(
  \nabla^i A_0 \nabla_i A_0 + \frac{e^{2 \sigma}}{2} A_0^2 f_{i j} f^{ij}  + \frac{e^{2 \sigma}}{2} F_{i j} F^{ij} 
 +e^{2 \sigma} A_0 f_{i j} F^{ij} \right) \frac{1}{T_0} Q^\prime(\nu)  \notag \\ 
& \quad  -\frac{1}{24\pi^2} A_0 \nabla^i \sigma \nabla_i \sigma + \frac{1}{8\pi^2} \nabla^i A_0 \nabla_i \sigma +
 \frac{1}{96\pi^2} e^{2\sigma} f_{ij} F^{ij} +  \frac{1}{32\pi^2} e^{2\sigma} A_0 f_{ij}f^{ij} +\frac{A_0}{48\pi^2} R   \,,
\end{align}
where $Q(\nu)$ is the analytic continuation of the series $Q(\nu) = -2\sum_{n=1}^\infty (-1)^n\cosh(n\nu)\log(n^2)$. A similar expression is obtained for $\langle T_{00}\rangle_2$. To derive this result we have regularized the vacuum contribution in a gauge invariant way by using the PV regularization procedure. We have defined the rescaled PV mass as~$\bar{M} = 2^{-3/2} e^{\gamma_E} M$. The distinction between vacuum and finite temperature and chemical potential contributions in the expectation values can be obtained, for instance, by considering the Poisson summation formula in the Matsubara sums.

\subsection{Partition function at second order}
\label{subsec:PF_2ndorder}

By using the variational formulae~(\ref{eq:varJ})-(\ref{eq:varT}) with Eq.~(\ref{eq:W2}), and after a comparison with the expressions of $\langle J_0\rangle_2$ and $\langle T_{00}\rangle_2$, one gets a system of 14 equations and 7 functions of two arguments. After consistently solving these equations we get the following result
\begin{align}
N_1(T,\nu) &= 0 = N_2(T,\nu) \,, \qquad M_1(T, \nu) = -\frac{1}{144} \frac{1}{T} - \frac{1}{48\pi^2} \frac{\nu^2}{T}  \,, \\
M_2(T, \nu) &=  \frac{1}{48\pi^2} T \left( \ln \frac{\bar{M}^2}{T^2} + Q(\nu) - 1 \right) \,,  \qquad M_3(T, \nu) = - \frac{1}{12\pi^2} \nu \,,  \\ 
M_4(T, \nu) &= -\frac{1}{96\pi^2} \frac{\nu^2}{T} \left( \ln \frac{\bar{M}^2}{T^2} + Q(\nu) + 3 - 6 \pi^2 C  \right) + \frac{1}{288}\frac{1}{T}  - \frac{C_2 }{8 T} + \frac{1}{384\pi^2} \frac{1}{T^3}  M^2 \ln 2 \,, \\
M_5(T, \nu) &=  -\frac{1}{96\pi^2} \frac{1}{T}  \left( \ln \frac{\bar{M}^2}{T^2} + Q(\nu)  \right) \,,  \qquad M_6(T, \nu) =  -\frac{1}{48\pi^2} \frac{\nu}{T} \left( \ln \frac{\bar{M}^2}{T^2} + Q(\nu) + 2 - 6 \pi^2 C   \right) \,,  \\
M_7(T, \nu) &= -\frac{1}{288} T - \frac{1}{96\pi^2} T \,  \nu^2  + \frac{1}{96 \pi^2} \frac{1}{T} M^2 \ln 2 \,,
\end{align}
where the constants $C$ and $C_2$ are given by Eq.~(\ref{eq:CC2}).  This result corresponds to a free theory of one left Weyl fermion. The coefficients obtained with one free Dirac fermion are twice these expressions. Note that these terms depend on the renormalization scale. The combination of terms proportional to $M^2$ in $M_4$ and $M_7$ is a pure renormalization effect, and they can be renormalized by adding a counterterm proportional to the Ricci scalar $\tilde{R}$ of the $3+1$ dimensional metric,
\begin{equation}
\cW_2^{\textrm{ct}} =  -\frac{M^2 \ln 2}{96 \pi^2}  \int d^4x \sqrt{-G} \,  \tilde{R} \,.
\end{equation}
The renormalized partition function is $\cW_2^{\textrm{ren}} = \cW_2 + \cW_2^{\textrm{ct}}$ and the counterterm exactly cancels the $M^2$ terms in $\cW_2$, so that the renormalized coefficients $M_{4,7}^{\textrm{ren}}$ are the same as $M_{4,7}$ after removing these contributions. On the other hand, this theory only violates conformal invariance because renormalization effects, which lead to a logarithmic dependence $\sim \ln \frac{\bar{M}}{T}$. The anomalous partition function reads
\begin{equation}
\begin{split}
\mathcal{W}_\mathrm{anom} &= \frac{1}{24 \pi^2} \int d^3 x \sqrt{g_3} \frac{1}{T}  \ln \frac{\bar{M}}{T} \times \left( e^{-2 \sigma}  g^{i j} \partial_i A_0 \partial_j A_0 - \frac{1}{2} A_0^2 f_{i j} f^{ij}  
  -\frac{1}{2} F_{i j} F^{ij} - A_0 f_{i j} F^{ij}\right)  \, \\
 &=- \frac{1}{48\pi^2} \int d^4 x  \sqrt{-G}
\ln \frac{\bar{M}}{T}  
\, \mathcal{F}_{\mu \nu} \mathcal{F}^{\mu \nu}   \,,
 \end{split}
\end{equation}
where $\mathcal{F}_{\mu\nu}$ is the four dimensional field strength of the gauge field~$\cA_\mu$. This is in agreement with the form of the local covariant action for the trace anomaly~\cite{Giannotti:2008cv,Eling:2013bj}.

\section{Non-dissipative constitutive relations}
\label{sec:const_rel}

In this section we determine partially the non-dissipative part of the second order constitutive relations in terms of the functions $M_i(T,\nu)$. The outline of the procedure may be sketched by
\begin{equation}
\langle \mathcal{O}_\ell \rangle_\mathrm{eq} = 
\delta (\mathcal{O}_\mathrm{perfect \,  fluid} +\mathcal{O}_{1} + \ldots + \mathcal{O}_{\ell-1} )   +  \mathcal{O}_\ell \,, 
\end{equation}
where $\mathcal{O}_\ell \equiv T_{(\ell)}^{\mu\nu}\,, \; J_{(\ell)}^\mu$, and $\delta (\mathcal{O}_\mathrm{perfect \,  fluid} + \cdots)$ is a correction of order $\ell$ due to all changes proportional to derivatives of the background evaluated in the constitutive relations of lower orders, see e.g. Ref.~\cite{Bhattacharyya:2013ida}. As an example $\delta \mathcal{O}_\mathrm{perfect \, fluid}$ receives corrections of the fluid velocity $\propto \delta u_{(1)}$ where $u = u_{(0)} + \delta u_{(1)} + \cdots$. The most general non-dissipative form of the constitutive relations in the Landau frame at second order can be expressed as
\begin{equation}
\begin{split}
T_{(2) \, \mu \nu} &=  T \left(\kappa_1 \tilde{R}_{\langle \mu \nu \rangle} + \kappa_2 u^{\alpha}  u^{\beta} \tilde{R}_{\langle \mu \alpha \nu\rangle \beta} + \kappa_3  \nabla_{\langle \mu} \nabla_{\nu \rangle} \nu  + \lambda_3 \,  \omega_{\langle \mu \alpha} \omega^{\alpha}_{\;\;\; \nu \rangle}  + \lambda_4 \, \mathfrak{a}_{\langle \mu} \mathfrak{a}_{\nu \rangle}   \right)  + \cdots  \\  
J_{(2) \, \mu} &= \upsilon_1 P_{\mu \alpha}  u_\nu \tilde{R}^{\nu \alpha}  + 
   \upsilon_2 P_{\mu \alpha} \nabla_\nu \mathcal{F}^{\nu \alpha}  + \cdots \,,
 \end{split}
\end{equation}
where~$ \mathfrak{a}_{\mu} = u^\alpha \nabla_\alpha u_\mu$, the vorticity tensor is $\omega_{\mu \nu} \equiv \frac{1}{2} P_{\mu}^\alpha P_{\nu}^\beta \left( \nabla_\alpha u_\beta - \nabla_\beta u_\alpha \right)$ and $X_{\langle \mu \nu \rangle}$ stands for the traceless and symmetric projection transverse to $u^\mu$, see e.g.~\cite{Kharzeev:2011ds,Banerjee:2012iz,Bhattacharyya:2013ida}. The goal is to determine the coefficients $\kappa_i$, $\lambda_i$ and $\upsilon_i$ by comparison with the partition function $\mathcal{W}_2$. After using the variational formulae Eqs.~(\ref{eq:varJ})-(\ref{eq:varT}) with Eq.~(\ref{eq:W2}), we arrive at the following general result
\begin{align}
&\kappa_1 = -2 M_7^{\textrm{ren}} \,, \qquad  \kappa_2 = -2 M_7^{\textrm{ren}} - 2 T \frac{\partial M_7^{\textrm{ren}}}{\partial T} \,, \qquad \kappa_3 = 2 \frac{\partial M_7^{\textrm{ren}}}{\partial \nu} \,, \\
&\lambda_3 |_{\mu = 0} = 16 T^2 M_4^{\textrm{ren}} - 6 M_7^{\textrm{ren}} - 2 T M_7^{\textrm{ren}\, \prime}(T) \,, \\
&\lambda_4 |_{\mu = 0} = -2 T^2 M_1 + 4 T M_7^{\textrm{ren}\, \prime}(T) + 2 T^2 M_7^{\textrm{ren}\, \prime\prime}(T) \,, \\
&\upsilon_1 = 4 T^2 \left(2 \nu M_5 - M_6 \right) - \frac{8\rho}{\varepsilon + P} T^3 
 \left(M_4^{\textrm{ren}} + \nu^2 M_5   - \nu  M_6  \right) \,,  \\
&\upsilon_2 =  - 4 T M_5  + \frac{2\rho}{\varepsilon + P} T^2 \left(2 \nu M_5 - M_6 \right) \,.
\end{align} 
The coefficients $\lambda_3$ and $\lambda_4$ multiply terms that are bilinear in derivatives, and these are more difficult to obtain. However they can be directly inferred at zero chemical potential from the results derived in Section 5 of Ref.~\cite{Banerjee:2012iz}. In the particular case of the free field theory of Weyl fermions, the explicit result of these transport coefficients are
\begin{align}
&\kappa_1 = \frac{T}{144} + \frac{1}{48\pi^2}\frac{\mu^2}{T} \,, \qquad \kappa_2 = 2 \kappa_1  \,, \qquad \kappa_3 =  -\frac{\mu}{24\pi^2}  \,, \qquad \lambda_3 |_{\mu = 0}  = 0 = \lambda_4 |_{\mu = 0} \,, \\
&\upsilon_1 =  -\frac{1}{2}\left(C - \frac{1}{3\pi^2} \right)\mu + \frac{\rho}{\varepsilon+P} \left[  \frac{1}{2}\left(C  - \frac{1}{6\pi^2}\right)\mu^2  + \left(C_2 - \frac{1}{36} \right)T^2  \right] \,, \\
&\upsilon_2 =  \frac{1}{24\pi^2} \left( \ln \frac{\bar{M}^2}{T^2}  + Q\left(\frac{\mu}{T}\right) \right)  - \frac{\rho}{\varepsilon + P} \frac{1}{4}\left(C - \frac{1}{3\pi^2} \right)\mu \,.
\end{align} 
The only second order coefficient which shows sensitivity to the renormalization scale is $\upsilon_2$. Notice also that there is no mixture between the conformal and chiral anomaly contributions in the constitutive relations. 

Let us compare these transport coefficients with some existing results in the literature. The values of $\kappa_{1,2}$ and $\lambda_3$ have been computed in Ref.~\cite{Moore:2012tc} at zero chemical potential. Our values for $\kappa_1$ and $\kappa_2$ are in agreement with this reference after setting $\mu=0$, but the vanishing value of $\lambda_3|_{\mu=0}$ is in contrast with the result obtained in this work, where they find~$\lambda_3^{\textrm{Moore,Sohrabi}} = -T^2/24$ for a Weyl fermion. This difference is related to the second term in the rhs of Eq.~(\ref{eq:Tp}), which seems to be not included in the diagrammatic computation of these authors. Note however that this coefficient was previously computed in Ref.~\cite{York:2008rr} from kinetic theory by some of the same authors, and a vanishing value was obtained. Finally the coefficients $\kappa_3$ and $\upsilon_2$ were explicitly computed within a holographic model in 5 dim in Refs.~\cite{Erdmenger:2008rm,Banerjee:2008th,Megias:2013joa}. A comparison between the weak coupling results presented above and the explicit expressions from the holographic model, leads to
\begin{align}
[\kappa_3]_{\textrm{weak coupling}} \propto [\kappa_3]_{\textrm{strong coupling}}  \,, \qquad [\upsilon_2]_{\textrm{weak coupling}} \sim c(T) + \frac{5}{112\pi^4}\frac{\mu^2}{T^2}  \propto [\upsilon_2]_{\textrm{strong coupling}}  \,,
\end{align}
in the regime $\mu \ll T$, where $c(T)$ has a logarithmic dependence on $T$ in the free fermion theory, while it is a constant in the holographic result. These coefficients receive contributions not induced by chiral anomalies, and so we cannot expect that the results from both approaches agree. However, we do confirm agreement in the parametric dependence in $\mu$ and $T$.

\section{Conclusions and discussion}
\label{sec:conclusions}

In this work we have studied second order transport effects induced by external electromagnetic fields, vortices and curvature, in a relativistic fluid for a free theory of chiral fermions in (3+1) dim. We have addressed the computation by using the equilibrium partition function formalism, which can only account for non-dissipative effects, i.e. transport coefficients multiplying quantities that survive in equilibrium. Examples of such contributions at first order in the hydrodynamic expansion are the chiral magnetic and vortical conductivities, which are $\cP$-odd and $\cT$-even. The situation is slightly different at second order, as we find that the parity violating part of the partition function vanishes, and the nonzero non-dissipative coefficients we have obtained are $\cP$-even and $\cT$-even. An explanation for this is that the only possible pseudo-scalars appearing at second order are $\cT$-odd, and so if the underlying Hamiltonian is invariant under $\cT$, this part of the partition function should vanish~\cite{Megias:2014mba}.

We have shown that the renormalization effects of the conformal anomaly mix with the chiral anomaly in some terms of the partition function, however this mixture does not appear in the constitutive relations. We examined the non-dissipative constitutive relations in the Landau frame, and derived the dependence  with temperature and chemical potential of five transport coefficients: $\kappa_{1,2,3}$ and $\upsilon_{1,2}$. We also obtained the zero chemical potential value for two additional coefficients $\lambda_{3,4}$. Some of them: $\kappa_{1,2,3}$, $\upsilon_2$ and $\lambda_3$; were previously computed in the literature by using different methods, and our results are consistent except for the latter. Finally let us mention that, as it has been pointed out in Refs.~\cite{Banerjee:2012iz,Megias:2014mba}, the vanishing value of  $\lambda_4$ at zero chemical potential is required by conformal invariance, while this only requires that  $\lambda_3 = \text{cons} \cdot T$. The vanishing value of  $\lambda_3$ is due to a further~cancellation.

\begin{acknowledgement}
This work has been supported by Plan Nacional de Altas Energ\'{\i}as grant FPA2012-34456, Spanish Consolider-Ingenio 2010 Programme CPAN (CSD2007-00042) and by the Basque Government under grant IT559-10. The research of E.M. is supported by the European Union under a Marie Curie Intra-European Fellowship (FP7-PEOPLE-2013-IEF) with project number PIEF-GA-2013-623006.
\end{acknowledgement}


%
%
%


\end{document}